\documentclass[twocolumn]{aastex701}

\usepackage{amsmath}
\usepackage{bm}
\usepackage{booktabs}
\usepackage{graphicx}
\usepackage{xcolor}
\newcommand{\ush}{u_{\mathrm{sh}}}
\newcommand{\fKH}{f_{\mathrm{KH}}}
\newcommand{\epsrec}{\epsilon_{\mathrm{rec}}}

\shorttitle{Photospheric KHI and Coronal Heating}
\shortauthors{Nykyri}

\begin{document}

\title{Photospheric Kelvin–Helmholtz Vortices as Possible Drivers of Coronal Heating: Implications of the DKIST Observations}

\author{Katariina Nykyri}
\affiliation{Department of Physical Sciences, Embry-Riddle Aeronautical University,
Daytona Beach, FL 32114, USA}
\affiliation{Center for Space and Atmospheric Research, Embry-Riddle Aeronautical University,
Daytona Beach, FL 32114, USA}
\email{nykyrik@erau.edu}

\begin{abstract}
The Daniel K. Inouye Solar Telescope (DKIST) has resolved Kelvin--Helmholtz
(KH) vortices at photospheric magnetic-flux boundaries with a characteristic
wavelength of 65 km.  I estimate whether these vortices can supply the
photospheric driver for cross-scale plasma heating through reconnection across different heights from photosphere to low-corona.  
Using the simulated MURaM shear, density contrast, and 500 km vertical extent, together with a representative
photospheric density, gives a shear-energy density of
$1.35\times10^{2}$ J m$^{-3}$ and $2.2\times10^{24}$ erg per characteristic
vortex.  Magnetic fields $1^\circ$--$7^\circ$ from the exact perpendicular orientation ($\bm B\perp\bm k$) remain KH unstable
in an idealized calculation and provide an in-plane component that can be wound
or compressed into current layers.  The limiting case, in which the
center-of-momentum shear reservoir becomes new magnetic free energy,  gives
$b_{\rm cs}=184$ G, identical to the ideal marginal-stability field and
equivalent to a $7.6^\circ$ effective twist.  This stores at most
135 J m$^{-3}$ in the layers.  Using empirical collisionless reconnection heating fractions of 0.28--0.44,
the same twist mapped to weakly collisional heights gives ion heating from
$\approx$20 eV at the photosphere to $\approx$1.4 keV in the low corona.  For
an illustrative, snapshot-based KH-active surface fraction of 0.03, quiet-Sun and coronal-hole losses
require 5--8\% and 14--21\%, respectively, of the shear reservoir to become
reconnecting magnetic free energy that reaches such heights.  Active regions
likely require a separate guide-field twist and helicity reservoir.  The
required upward transport has not been measured by DKIST, but it is directly
testable.
\end{abstract}


\section{Introduction}\label{sec:intro}

The temperature rises from about 5800 K at the visible solar surface to more
than $10^6$ K in the corona.  The photosphere occupies approximately the first
0.5 Mm above the mean optical-depth-unity surface, the chromosphere extends to
roughly 2 Mm, and the transition region separates it from coronal-temperature
plasma.  These heights vary with the magnetic and thermodynamic structure
\citep{avrett2008}.  Representative energy losses are about
300 W m$^{-2}$ for the quiet-Sun corona, 800 W m$^{-2}$ in coronal holes, and
$10^4$ W m$^{-2}$ in active regions \citep{withbroe1977,cranmer2019}.  Coronal
heating is usually discussed in terms of magnetic stressing and reconnection,
or the dissipation of waves and turbulence, although the two routes need not
remain separate once nonlinear structure develops
\citep{klimchuk2006,klimchuk2015,parnell2012,demoortel2015,
vandoorsselaere2020}.  A process
driven below or within the photosphere must satisfy two conditions:
it must contain sufficient free energy, and part of that energy must cross the
chromosphere.  The Kelvin-Helmholtz Instability (KHI) is relevant to this problem because it can turn an
observable shear flow into both thin current layers, shorter-wavelength fluctuations and plasma waves enabling ion and electron heating \citep{masson2018,nykyriMaJohnson2021} .

Photospheric footpoint motions provide a direct magnetic route into this
problem.  As emphasized by \citet{klimchuk2015}, the observed motions
continually tangle and twist the coronal field; reconnection is then required
to prevent magnetic stress from increasing without bound.  In the Parker
nanoflare picture, this stressing produces many thin current layers whose
intermittent reconnection releases the stored free energy
\citep{parker1988,klimchuk2006,wilmotSmith2015}.  Numerical experiments show
that random footpoint driving, magnetic braiding, and helicity redistribution
can sustain a hierarchy of current sheets and impulsive heating events
\citep{rappazzo2008,knizhnik2019,klimchuk2023}.  Solar Orbiter observations
have also associated the relaxation of small-scale coronal braids with
localized impulsive heating \citep{chitta2022}.  The KH mechanism considered
here belongs within this broader picture: a photospheric KH vortex is a
resolved, localized footpoint motion that may help build and compress the
current layers in which energy is released.

The heating calculations in this Letter are based on energy budget estimation using  DKIST observations and the recently obtained scaling laws of collisionless magnetic reconnection. However, they are not yet a proof that the DKIST observed vortices
already heat the corona.  The controlling unknown is the product of the
shear-to-magnetic energy conversion and the upward survival of that stress through the
chromosphere into corona.

\citet{kuridze2026} have identified a new source of small-scale photospheric
shear.  DKIST continuum images at 19 km resolution reveal
KH-unstable interfaces around magnetic flux concentrations, with 47 vortices
analyzed in the observed field.  The measured wavelengths
span 25--170 km and peak near 65 km.  Apparent speeds are 0.67--3.0 km s$^{-1}$,
and measured linear growth rates are 0.014--0.054 s$^{-1}$.  A companion MURaM
radiation-MHD calculation reproduces the structures and shows a shear layer
approximately 12 km wide.  The KH rolls remain coherent from about 100 km above
to 400 km below the mean $\tau_{500}=1$ surface.  The observations establish
photospheric KHI and mixing; they do not measure reconnection, upward Poynting
flux, or coronal heating. 

Solar KHI has been observed or modeled at CME boundaries, in the extended
corona and solar wind, and in jets and prominence flows
\citep{ofman2011,foullon2011,foullon2013,nykyriFoullon2013,zaqarashvili2015,kuridze2016,
li2018,hillier2018,paouris2024,Nykyri2024,ofman2026}.  These events show that shear
instability is common in the solar atmosphere and its transients.

Solar simulations provide the more relevant link.  Photospheric twisting can
propagate upward in a stratified flux tube \citep{murawski2016}; oscillating
coronal loops are KH unstable at their boundaries
\citep{terradas2008,soler2010}; and nonlinear loop calculations produce fine
current layers, turbulent mixing, reconnection, and enhanced dissipation
\citep{antolin2014,howson2017,karampelas2017,howson2021,shi2021}.  These
calculations use drivers and plasma descriptions different from the DKIST
event, but they establish that KH-assisted current formation is viable in
coronal plasma.

In situ magnetospheric studies add the kinetic-scale part of the sequence that
cannot yet be resolved in the corona.  Simulations and spacecraft observations
show nonlinear KH vortices forming thin current sheets, multiple reconnection
sites, turbulence, ion-scale waves, and particle heating
\citep{otto2000,nykyri2001,nykyri2004,nakamura2006,nykyri2006,
eriksson2016,li2016,stawarz2016,vernisse2016,wilder2016,nakamura2017,
moore2016,moore2017,hasegawa2020}.  Three-dimensional simulations also produce
paired reconnection sites and twice-reconnected flux \citep{otto2008,faganello2012,ma2017,faganello2017}.  I use
these magnetospheric results to identify candidate processes and measured energy partitions, that could be further tested in future DKIST motivated high-resolution simulations coupling photosphere to corona. I cannot yet assign a solar conversion efficiency, because the lower solar atmosphere
is partially ionized, stratified, and more collisional than the magentosheath-magnetopause-magnetosphere system.

\section{Observational Constraints and Magnetic Geometry}\label{sec:model}

Table~\ref{tab:inputs} separates the DKIST measurements from quantities derived
from MURaM and from assumptions introduced below.  This distinction is
important because the present DKIST data are continuum intensities rather than
co-spatial vector magnetic-field and velocity measurements.

\begin{table*}[t]
\centering
\caption{Quantities used in the estimate.\label{tab:inputs}}
\footnotesize
\setlength{\tabcolsep}{4.5pt}
\begin{tabular}{llll}
\toprule
Quantity & Value & Status & Use \\
\midrule
KH vortices analyzed & 47 & DKIST measurement & observed sample \\
Wavelength & 65 km (25--170 km) & DKIST measurement & characteristic footprint \\
Growth rate & 0.014--0.054 s$^{-1}$ (prominent examples) & DKIST measurement & observed time scale \\
Apparent speed & 0.67--3.0 km s$^{-1}$ & DKIST measurement & pattern propagation \\
Shear width, $\Delta U$ & 12 km, 3.0 km s$^{-1}$ & MURaM analysis & shear reservoir \\
Roll depth, density contrast & 500 km, $\rho_1/\rho_2=4$ & MURaM analysis & source volume and asymmetry \\
$B_z$, $\rho_1$ & 1.4 kG, $3.0\times10^{-4}$ kg m$^{-3}$ & representative normalization & angle and energy estimate \\
$\alpha$ & $1^\circ$--$7^\circ$ & sensitivity range & in-plane field component \\
$T$ & 150 s & illustrative KH evolution/renewal time & areal heat-flux normalization \\
$\fKH$ & 0.03 & MURaM snapshot estimate & mean KH-active fraction \\
\bottomrule
\end{tabular}
\end{table*}

The fiducial $T=150$~s is a rounded characteristic KH evolution and
energy-renewal time, not a measured recurrence period.  It is comparable to
both the approximately 180~s DKIST observing sequence and the approximately
174~s interval over which the MURaM calculation shows nonlinear vortex
evolution \citep{kuridze2026}.  It also corresponds to 2.1--8.1 e-folding
times for the measured growth-rate range, since
$\gamma^{-1}=18.5$--71.4~s.  I define $\fKH$ as the spatially and temporally
averaged surface fraction occupied by active KH structures.  If the occurrence
statistics are separated explicitly, $\fKH=f_{\rm site}f_{\rm duty}$, where
$f_{\rm site}$ is the surface fraction capable of hosting KHI and
$f_{\rm duty}=\tau_{\rm active}/T_{\rm rec}$ is the fraction of time those
sites are active.  The adopted value $\fKH=0.03$ instead comes directly from
the instantaneous active-area estimate
$94(3\lambda)\lambda/(6.144\,{\rm Mm})^2=0.0316$ in one MURaM synthetic
snapshot.  The 94 interfaces are counted in the synthetic image, whose
wavelength histogram peaks at 49 km rather than the observed 65 km; the same
footprint estimate with 49 km gives 0.018, within a factor of two of the
adopted normalization.  Its use in the mean heat flux assumes that this snapshot is
representative of the time average; it does not constitute a separate
measurement of $f_{\rm duty}$.

Let $\bm k$ be directed along the horizontal velocity shear.  If the field is
tilted by $\theta$ from the surface normal and its horizontal projection makes
an angle $\varphi$ with $\bm k$, the component available for in-plane tension
and reconnection is
\begin{equation}
 B_{\parallel}=B\sin\theta\cos\varphi\equiv B\sin\alpha .
 \label{eq:bparallel}
\end{equation}
Thus, $\alpha$ is the effective departure from $\bm B\perp\bm k$; it is not the
field inclination alone.  For $B=1.4$ kG, $\alpha=1^\circ$, $3^\circ$,
$5^\circ$, and $7^\circ$ gives $B_{\parallel}=24$, 73, 122, and 171 G.  The
corresponding magnetic-energy densities in parallel fields, $u_B=B_{\parallel}^2/(2\mu_0)$,  are thus 2.4, 21, 59, and
116 J m$^{-3}$.  This is pre-existing magnetic energy which KHI can further rearrange and
compress, but only the magnetic-energy increment produced by the vortex can
be counted as conversion of shear energy.

  Magnetic tension from $B_{\parallel}$ also reduces the growth rate.  For an
incompressible tangential discontinuity the instability requires
\begin{equation}
 [\bm k\!\cdot\!(\bm U_1-\bm U_2)]^2>\frac{1}{\mu_0}
 \left(\frac{1}{\rho_1}+\frac{1}{\rho_2}\right)
 \left[(\bm k\!\cdot\!\bm B_1)^2+(\bm k\!\cdot\!\bm B_2)^2\right]
 \label{eq:chandra}
\end{equation}
\citep{chandrasekhar1961}.  Taking $\bm k\parallel\Delta\bm U$ and equal in-plane
components on both sides, $\bm k\!\cdot\!\bm B_{1,2}=kB_{\parallel}$, the factor
$k^2$ cancels and 
marginal stability occurs
at
\begin{equation}
 B_{\parallel,c}=\Delta U
 \left[\frac{\mu_0\rho_1\rho_2}{2(\rho_1+\rho_2)}\right]^{1/2}
 =184\ {\rm G} .
 \label{eq:bcrit}
\end{equation}
The corresponding angle is $\alpha_c=7.56^\circ$, and
\begin{equation}
 \frac{\gamma}{\gamma_0}=
 \left[1-\left(\frac{B_{\parallel}}{B_{\parallel,c}}\right)^2\right]^{1/2}.
 \label{eq:growth}
\end{equation}
This gives 0.99, 0.92, 0.75, and 0.38 as KHI growth-rate ratio between the finite $B_{\parallel}$-cases and the idealized, singular ${\bf{k}}\perp {\bf{B}}$ -case ($\gamma_0$) for the four angles above.  This provides an angular sensitivity estimation which can be checked in future MURaM simulations and DKIST observations: finite shear layer width, compressibility,
stratification, and the weakly magnetized granular side modify the threshold.
Two-fluid calculations for partially ionized solar flux tubes also show that
neutrals and ion--neutral collisions can change the ideal-MHD onset and growth
rates \citep{martinezGomez2015}.  The estimate is nevertheless consistent with
the nearly perpendicular fields required for the observed photospheric KHI and
with the small field tilts found for the KHI unstable CME boundary at low corona by
\citet{nykyriFoullon2013}.

\begin{figure*}[t]
\centering
\includegraphics[width=0.98\textwidth]{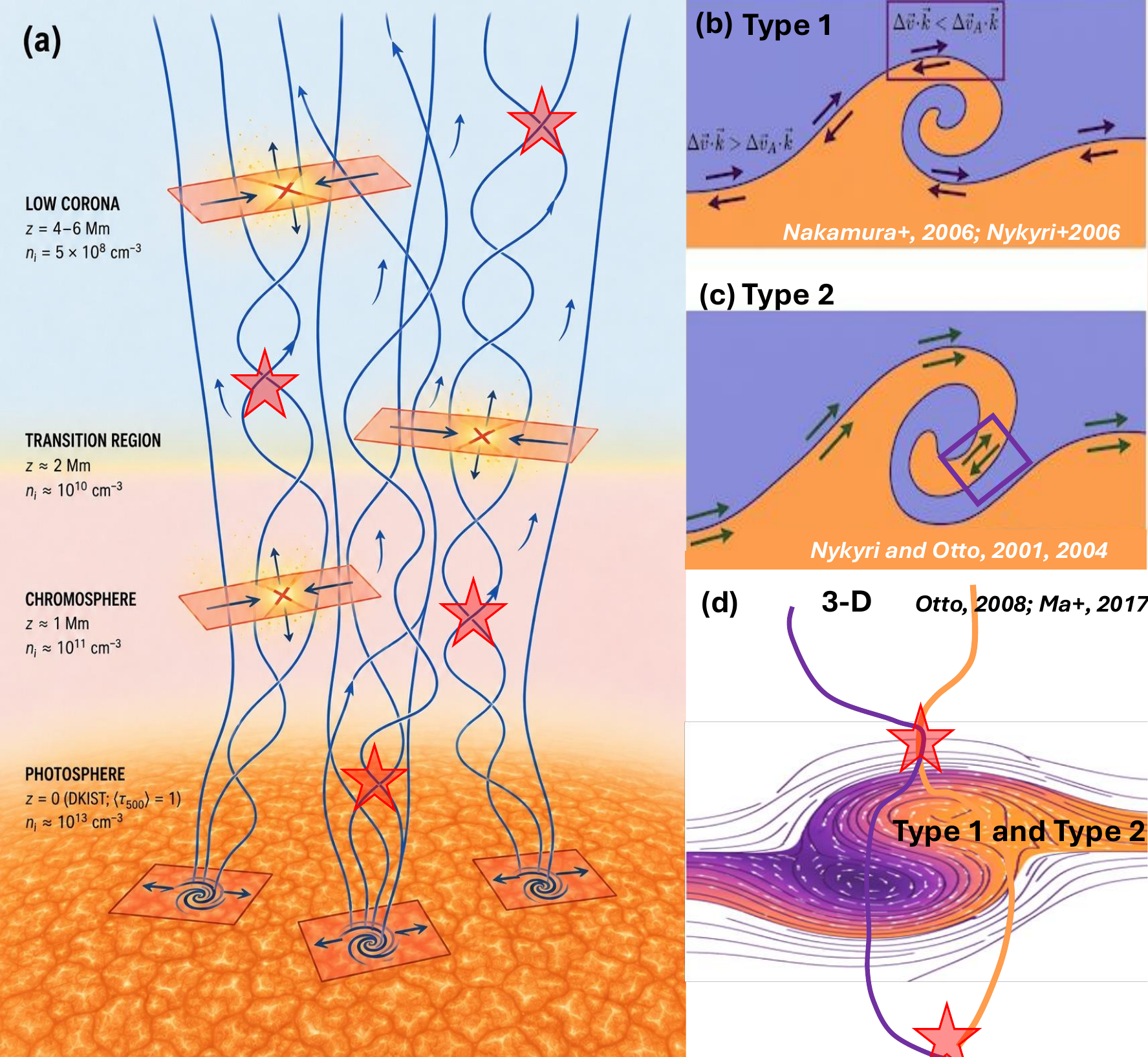}
\centering
\caption{A possible cross-scale path from photospheric KH forcing to the low
corona through KH driven reconnection and component reconnection. (a): A photospheric DKIST-scale tornado-like vortex acts at the footpoint of an expanding twisted magnetic structure at the
magnetic-flux boundaries, with short horizontal, reconnecting current layers at several
heights, locally perpendicular to the dominant guide field that extends from photosphere to corona.
(b)--(c): local 2-D KHI driven reconnection paths for pre-existing
anti-parallel in-plane fields (Type 1) \cite{nakamura2006,nykyri2006} and for initially parallel in-plane
fields that are wound by KH vortex motion into compressed current layers (Type 2) \cite{nykyri2001,nykyri2004}.
(d): 3-D paired "candy-wrapper" reconnection and
twice-reconnected flux below and above the shear-flow-plane \citep{otto2008,faganello2012,ma2017,faganello2017}. The star symbol marks the location of three-dimensional magnetic reconnection in panels (a) and (d). In 3-D  all the mechanisms can occur: The KH vortex driven Type 1 and Type 2 reconnection take place in the shear-flow-plane. Component reconnection can also occur in the absence of any flow shear in the plane perpendicular to main magnetic field. The fields perpendicular to the shear flow plane can be either parallel or anti-parallel.}
\label{fig:KHI}
\vspace{0.5em}
\end{figure*}

Figure~\ref{fig:KHI}a places the reconnection geometries in the proposed
photosphere-to-corona sequence.  Figure~\ref{fig:KHI}b-d show the possible 2-D and 3-D reconnection geometries enabled by photospheric KHI dynamics. 
In the 2-D geometry (Type 2, panel c) , initially parallel
in-plane fields are wound by the
vortex until adjacent turns become antiparallel.  The simulations of
\citet{nykyri2001} produced multiple filamentary current layers and magnetic
islands in this manner.  Hall-MHD calculations retained a mass-transport speed
of order 1.5 km s$^{-1}$ and generated structures near the ion inertial length
\citep{nykyri2004}.  A second 2-D geometry (Type 1, panel b)  begins with an antiparallel magnetic fields across the current
sheet, which the KH flow compresses and brings to reconnection \citep{nakamura2006,nykyri2006}.  Solar-coronal calculations likewise find that KHI can
promote reconnection and that reconnection occurs preferentially along vortex
boundaries \citep{lapenta2003,howson2021}.  In three dimensions, KHI driven out-of-the shear-flow plane twisting of the magnetic field can
produce reconnection sites both above and below the central vortex plane even when the fields are initially parallel. This geometry also works for anti-parallel geometry where KHI produces fast reconnection rate even in the absence of the Hall-term \citep{mao2014a,mao2014b}.  Field
lines that pass through both sites become twice reconnected \citep{otto2008,faganello2012,ma2017,faganello2017} resulting in strong mass transport;
spacecraft observations have independently resolved multiple reconnection
sites along KH-unstable boundaries
\citep{nykyri2006,eriksson2016,li2016,nakamura2017,li2023} both for northward
and southward IMF.  Along a solar flux tube, these processes need not occur at
one height.  The photospheric KH roll can act as a vortex-like footpoint driver
while thin current layers form at several altitudes, locally perpendicular to
the guide field and reconnecting wherever the plasma and field geometry permit: For 2-D reconnection the difference between Alfv\'en speed along KHI k-vector needs to be greater than the velocity shear along the k-vector ($\Delta V_A \cdot k > \Delta V \cdot k$)), whereas the 3-D reconnection is initiated when the intense twisting cuts-off the field aligned current.
Figure~\ref{fig:KHI} does not imply a single X-line extending from the
photosphere into the corona, but rather several patchy reconnection sites where
magnetic energy, amplified by the vortex motion, is transferred to heat and to
kinetic energy of the particles.

The two 2-D paths require different amounts of vortex winding.  If an
anti-parallel in-plane component is already present, the KH flow mainly
compresses and thins the sheet.  If the in-plane fields are initially parallel,
the vortex must wind the field until adjacent turns become locally
anti-parallel.  Anti-parallel out-of-plane components can occur in more complex
magnetic topologies, but they are not required for the quiet-Sun 2-D heating
estimate.

\section{Energy and Heating Estimates}\label{sec:energy}

\subsection{Shear energy in one vortex}

The relevant kinetic energy of the KH wave is that of the relative motion of the adjacent plasma flows.  For two
equal-volume layers of plasmas with different densities, the center-of-momentum shear-energy density is
\begin{equation}
 \ush=\frac{1}{4}\frac{\rho_1\rho_2}{\rho_1+\rho_2}(\Delta U)^2 .
 \label{eq:ush}
\end{equation}
The density contrast $\rho_1/\rho_2=4$ is the mean MURaM value used in the
linear analysis of \citet{kuridze2026}, tabulated there as
$\epsilon\equiv(\rho_1-\rho_2)/(\rho_1+\rho_2)=0.6$ in their Extended Data Figure 6e and 
Table~1.

For the absolute normalization, I adopt
$\rho_1=3.0\times10^{-4}$ kg m$^{-3}$, close to the
$(2.6$--$2.8)\times10^{-4}$ kg m$^{-3}$ implied by the
quiet-Sun Model C atmosphere near $\tau_{500}=1$
\citep{fontenla2006,avrett2008}. This density is an adopted
reference value rather than a quantity measured in the DKIST event.

 With
$\Delta U=3.0$ km s$^{-1}$, Equation~(\ref{eq:ush}) gives
$\ush=135$ J m$^{-3}$.  Treating the 65 km wavelength as a characteristic
vortex diameter gives
\begin{equation}
 E_{\mathrm{sh}}=\ush\,\pi(\lambda/2)^2 H
 =2.24\times10^{17}\ {\rm J}=2.24\times10^{24}\ {\rm erg},
 \label{eq:evortex}
\end{equation}
for $H=500$ km.  Since density and shear vary with height, this is an
order-of-magnitude column estimate.  Its dependence is explicit:
$E_{\mathrm{sh}}\propto\rho_{\rm red}(\Delta U)^2\lambda^2H$.

Let $\xi_B$ be the fraction of this shear reservoir converted into magnetic
free energy in thin current sheets.  Energy conservation requires
\begin{equation}
 \Delta u_B=\xi_B\ush,\qquad 0\leq\xi_B\leq1 .
 \label{eq:xib}
\end{equation}
Two-dimensional and three-dimensional reconnection can both occur within a KH
wave, but their released energies are parts of the same $\Delta u_B$.

\subsection{From shear flow to thin current layers}
\label{sec:layers}

The limiting case $\xi_B=1$ fixes the largest field increment that vortex
winding and compression can build in the layers.  Setting $\Delta u_B=\ush$ in
Equation~(\ref{eq:xib}) and writing $\Delta u_B=b_{\rm cs}^2/(2\mu_0)$ for the
amplified in-plane component $b_{\rm cs}$ gives
\begin{equation}
 b_{\rm cs}=\left(2\mu_0\ush\right)^{1/2}
 =\Delta U\left[\frac{\mu_0\rho_1\rho_2}{2(\rho_1+\rho_2)}\right]^{1/2}
 =184\ {\rm G},
 \label{eq:bcs}
\end{equation}
which is \emph{identical} to the marginal-stability field
$B_{\parallel,c}$ of Equation~(\ref{eq:bcrit}).  The winding is then
self-limiting: as the wound component approaches $B_{\parallel,c}$, the
residual growth rate in Equation~(\ref{eq:growth}) vanishes and the driver
shuts off.  This is an upper bound, not an assumed efficiency.  Complete
conversion of the center-of-momentum shear energy would twist the field by at
most $\alpha_c=7.6^\circ$, comparable to, though somewhat
below, the $10^\circ$--$20^\circ$ Parker angle inferred from coronal
Poynting-flux arguments \citep{klimchuk2015}.  The stored magnetic energy
density in the layers is then
\begin{equation}
 u_{\rm cs}=\frac{b_{\rm cs}^2}{2\mu_0}=\xi_B\,\ush
 =135\ {\rm J\,m^{-3}}\quad(\xi_B=1),
 \label{eq:ucs}
\end{equation}
in addition to the 2.4--116 J m$^{-3}$ already present in the seed component
(Section~\ref{sec:model}), which the vortex rearranges but does not create.

Winding through $N$ turns interleaves oppositely directed layers with spacing
$\sim\lambda/(2\pi N)$ \citep{nykyri2001}, and the vortex-scale strain
compresses each layer.  In collisionless KH calculations, reconnection onsets
when the layer approaches the ion kinetic scale \citep{nykyri2004}.  The
corresponding inertial lengths are
\begin{equation}
 d_i=\frac{c}{\omega_{pi}}
 =\frac{2.28\times10^{7}}{\sqrt{n_i\,[{\rm cm^{-3}}]}}\ {\rm cm},
 \qquad d_e=\left(\frac{m_e}{m_i}\right)^{1/2}d_i .
 \label{eq:dide}
\end{equation}
For the photospheric ionized-component density $n_i=10^{13}$ cm$^{-3}$
(ionization fraction $n_i/n_{\rm H}\approx6\times10^{-5}$ for the adopted $\rho_1$),
$d_i=7.2$ cm and $d_e=1.7$ mm.  A layer of length $\sim\lambda$ compressed to
$d_i$ has an aspect ratio $\lambda/d_i\sim10^{6}$, far
beyond the onset conditions for the plasmoid instability of thin sheets,
whether expressed as a critical Lundquist number $S\gtrsim10^{4}$ for
Sweet--Parker layers \citep{loureiro2007} or as a critical aspect ratio of
order $10^{2}$ \citep{cassak2009}.  Such a sheet would be prone to bursty secondary
islanding if a sufficiently conducting ion--electron layer survives to this
aspect ratio.  Flux conservation during compression amplifies the local field
by the compression ratio, but total-pressure balance
caps the layer field near $(2\mu_0 p_{\rm ph})^{1/2}\simeq1.7$ kG for
$p_{\rm ph}\simeq1.2\times10^{4}$ Pa; the increment attributable to shear
energy saturates at Equation~(\ref{eq:bcs}).  The free energy per unit sheet
area at the ion width is $u_{\rm cs}d_i\simeq10$ J m$^{-2}$.  In the weakly
ionized photosphere these collisionless widths apply to the ion--electron
fluid only after the sheet decouples from the neutrals.  Two-fluid effects
modify the onset \citep{martinezGomez2015}, and simulations of weakly ionized
current sheets show fast reconnection once kinetic scales are approached
\citep{leake2012}.  The collisionless scaling is most secure for the
upper chromosphere, transition region, and low corona, and should be treated as
a photospheric upper limit until the ion--neutral coupling is measured or
modeled for the DKIST vortices.

\subsection{Reconnection heating}

Spacecraft observations give an empirical scale for the conversion after an
ion-scale current sheet has formed.  At the magnetopause, the electron and ion thermal energy increase scales as 
$k_{\rm B}\Delta T_e=0.017m_iV_A^2$ and
$k_{\rm B}\Delta T_i=0.13m_iV_A^2$
\citep{phan2013,phan2014}.  In high-$V_A$, low-$\beta$ magnetotail exhausts,
electron heating follows $k_{\rm B}\Delta T_e=0.020m_iV_A^2$
\citep{oieroset2023}, whereas ion heating becomes sublinear at the largest
$m_iV_A^2$ \citep{oieroset2024}.  A linear fit to the total ion-plus-electron
heating in the latter study is $0.093m_iV_A^2$.

For scalar bulk temperatures,
$\Delta u_{\rm th}=(3/2)nk_{\rm B}(\Delta T_i+\Delta T_e)$.  Since
$nm_iV_A^2=2u_B$, the magnetopause coefficients give
$\Delta u_{\rm th}/u_B=0.441$, while the high-$V_A$ total-heating fit gives
0.279.  I use
\begin{equation}
 \epsrec\equiv\frac{\Delta u_{\rm th}}{u_B}=0.28\text{--}0.44
 \label{eq:epsrec}
\end{equation}
as an empirical bracket.  It applies to local collisionless exhausts, not to
the partially ionized photosphere.  Combining Equations~(\ref{eq:ush})--
(\ref{eq:epsrec}) gives
\begin{align}
 \Delta u_{\rm th} &=38\text{--}60\,\xi_B\ {\rm J\,m^{-3}},\\
 E_{\rm th} &=(6.3\text{--}9.9)\times10^{23}\xi_B\ {\rm erg}
 \label{eq:localheat}
\end{align}
per characteristic source volume.  Nonthermal electron energization can also
occur.  The 20--60\% nonthermal energy fractions reported in selected
magnetotail reconnection intervals \citep{oka2022} describe the electron
population in those events and are not added to Equation~(\ref{eq:localheat}).
Three-dimensional acceleration by parallel electric fields, Fermi reflection,
and betatron processes remains a possible additional partition of the released
energy \citep{oka2023}.

\subsection{Heating across the atmosphere}
\label{sec:ladder}

The per-particle heating implied by these scalings depends on altitude only
through the local density and field strength.  If the equipartition twist of
Section~\ref{sec:layers} propagates along the expanding flux tube with its
angle approximately preserved---an assumption that forms part of
$\eta_{\rm up}$---the reconnecting component at height $h$ is
\begin{equation}
 b_{\rm cs}(h)=B(h)\sin\alpha_c,\qquad \sin\alpha_c=0.132,
 \label{eq:bh}
\end{equation}
and the inflow Alfv\'en speed based on the reconnecting component is
$V_{\!A,\rm cs}=b_{\rm cs}/(\mu_0 n_i m_i)^{1/2}$.
Table~\ref{tab:ladder} evaluates
Equations~(\ref{eq:dide})--(\ref{eq:bh}) and the magnetopause coefficients
$k_{\rm B}\Delta T_i=0.13\,m_iV_{\!A,\rm cs}^2$ and
$k_{\rm B}\Delta T_e=0.017\,m_iV_{\!A,\rm cs}^2$
\citep{phan2013,phan2014} for representative quiet-network parameters,
with $B(h)$ decreasing from the 1.4 kG footpoint to 80 G in the low corona
and $n_i$ from the quiet-Sun stratification
\citep{fontenla2006,avrett2008}.

\begin{table*}[t]
\centering
\caption{Current-layer properties and reconnection heating along the flux
tube for a preserved twist $\alpha_c=7.6^\circ$ and complete conversion
($\xi_B=1$).\label{tab:ladder}}
\footnotesize
\setlength{\tabcolsep}{5pt}
\begin{tabular}{lccccccccc}
\toprule
Region & $n_i$ & $B$ & $b_{\rm cs}$ & $d_i$ & $d_e$ &
$u_{\rm cs}$ & $V_{\!A,\rm cs}$ & $k_{\rm B}\Delta T_i$ &
$k_{\rm B}\Delta T_e$ \\
 & (cm$^{-3}$) & (G) & (G) & & & (J m$^{-3}$) & (km s$^{-1}$) & (eV) & (eV) \\
\midrule
Photosphere (0--0.5 Mm)     & $10^{13}$        & 1400 & 184 & 7.2 cm & 1.7 mm & 135  & 127  & 22   & 2.9 \\
Low chromosphere ($\sim$1 Mm)  & $10^{11}$     & 300  & 39  & 72 cm  & 1.7 cm & 6.2  & 272  & 101  & 13  \\
Upper chromosphere ($\sim$2 Mm) & $10^{10}$    & 150  & 20  & 2.3 m  & 5.3 cm & 1.5  & 430  & 252  & 33  \\
Low corona (4--6 Mm)        & $5\times10^{8}$ & 80   & 10.5 & 10 m   & 24 cm  & 0.44 & 1030 & 1430 & 187 \\
\bottomrule
\end{tabular}
\end{table*}

Although the stored energy density $u_{\rm cs}=b_{\rm cs}^2/(2\mu_0)$
decreases with height, the available energy per particle,
$m_iV_{\!A,\rm cs}^2=2u_{\rm cs}/n_i$, increases because the density falls
faster than $B^2$. Therefore, dissipation of these current sheets produces modest heating at low altitudes 
and increasingly energetic per-particle heating at greater heights.  At
the photosphere the exhaust ions gain $\approx$22 eV
($2.5\times10^{5}$ K); collisional sharing with the
$\sim1.8\times10^{4}$-fold more numerous neutrals subsequently dilutes the
bulk temperature rise to $\sim10^{-3}$ eV ($\sim$14 K) per event, so
photospheric sheets dissipate shear energy locally without producing hot
plasma.  Chromospheric sheets heat ions to $10^{2}$--$2.5\times10^{2}$ eV
($1$--$3\times10^{6}$ K) before neutral dilution.  In the low corona the same
footpoint twist yields $k_{\rm B}\Delta T_i\approx1.4$ keV
($1.7\times10^{7}$ K) and $k_{\rm B}\Delta T_e\approx0.19$ keV
($2.2\times10^{6}$ K); at this altitude
$m_iV_{\!A,\rm cs}^2\approx11$ keV lies in the high-$V_A$ regime where the
sublinear magnetotail fit is more appropriate, giving a combined
ion-plus-electron heating of $\approx$1.0 keV \citep{oieroset2024}.
The four values of $V_{\!A,\rm cs}$ in
Table~\ref{tab:ladder} (127--1030 km s$^{-1}$) all lie within the combined
Alfv\'en-speed range over which these empirical relations have been validated
(10--600 km s$^{-1}$ at the magnetopause and 800--4000 km s$^{-1}$ in the
magnetotail), a regime that \citet{oieroset2024} note could be applicable to
the solar corona and flare environments.  Both coefficients yield coronal
exhaust temperatures in the nanoflare regime, consistent with
the hot ($\sim10^{7}$ K) plasma components inferred in active-region cores
\citep{klimchuk2015}.

Two caveats accompany Table~\ref{tab:ladder}: the large
guide-to-reconnecting-field ratio,
$B_g/b_{\rm cs}=\cot\alpha_c\simeq 7.5$, can reduce the
ion-heating coefficient below its antiparallel-reconnection value;
the empirical coefficients were calibrated at guide-to-reconnecting ratios of
0--1 at the magnetopause \citep{phan2013,phan2014} and $\lesssim$0.2 in the
magnetotail \citep{oieroset2024}, so their use at
$B_g/b_{\rm cs}\simeq7.5$ is an extrapolation.  Meanwhile, the
free energy per unit sheet area at the ion inertial scale,
$u_{\rm cs}d_i\simeq 3$--$10\ {\rm J\,m^{-2}}$, is nearly independent of
height. Consequently, the number of particles processed by each sheet,
rather than the energy content per sheet, controls the altitude dependence.

The per-particle temperature increments in Table~\ref{tab:ladder} describe
individual reconnection events and are independent of their occurrence rate.
The occurrence statistics are carried only by the spatially and temporally
averaged active fraction $\fKH$ in the areal heat flux below.

The heat deposited directly in the photosphere does not determine the coronal
heating rate.  Instead, the photospheric KH vortices affect the corona
indirectly by twisting the magnetic field and transferring magnetic stress
upward along the expanding flux tube.  Coronal heating requires that a
fraction of this magnetic free energy reach sufficiently ionized atmospheric
layers before being dissipated or transferred to the largely neutral
photospheric plasma.

\subsection{Areal heat flux}

The mean coronal heat flux depends on $\fKH$ and on $\eta_{\rm up}$, the
fraction of the KH-generated magnetic free energy that is transported to a
sufficiently ionized layer.  Neither quantity has yet been constrained by
DKIST.  The resulting energy-flux bookkeeping is
$\ush\rightarrow\xi_B\ush\rightarrow\eta_{\rm up}\xi_B\ush
\rightarrow\epsrec\eta_{\rm up}\xi_B\ush$.  Here $\xi_B$ is the
shear-to-magnetic conversion, $\eta_{\rm up}$ is the survival and
upward-transport factor, and $\epsrec$ is the local reconnection heating
fraction.  The mean heat flux is then
\begin{align}
 F_{\rm heat}&=\fKH\,\xi_B\eta_{\rm up}\epsrec
               \frac{\ush H}{T} \notag\\
 &=1.35\times10^4\,\xi_B\eta_{\rm up}\epsrec
 \left(\frac{\fKH}{0.03}\right)
 \left(\frac{150\ {\rm s}}{T}\right)\ {\rm W\,m^{-2}} .
 \label{eq:flux}
\end{align}

The normalization $\fKH=0.03$ uses the instantaneous active-area fraction from
the MURaM snapshot described after Table~\ref{tab:inputs}, while $T=150$~s is
the adopted KH evolution and energy-renewal time.  Neither quantity is a
measured recurrence statistic, and the temporal duty cycle remains
unconstrained.

For this normalization, the quiet-Sun loss requires
$\xi_B\eta_{\rm up}=0.051$--0.079, and the coronal-hole loss requires
0.135--0.212, where the range reflects Equation~(\ref{eq:epsrec}).  The
active-region requirement gives 1.68--2.65 and is not satisfied.
These values show what the proposed mechanism would have to accomplish; they
are not measurements of its efficiency.

\section{Cross-Scale Wave Heating}\label{sec:waves}

The current sheets need not be the final dissipation scale.  Magnetospheric KH
events show fast/kinetic magnetosonic packets, KAW Poynting flux, intermittent
turbulence, and large-amplitude electrostatic waves inside or near KH vortices
\citep{moore2016,moore2017,stawarz2016,wilder2016,hasegawa2020,nykyri2021}.
These waves can carry a substantial part of the local shear-energy surplus and
can heat particles, but their solar efficiency is not known.

A direct scale mapping from the magnetosphere to the solar
photosphere and atmosphere is not sufficient by itself.  In the Cluster
spacecraft event, KHI-driven fast/kinetic magnetosonic waves (FMWs) produced
1--2~keV ion heating \citep{moore2016} and had
$\lambda_{\rm KH}/\lambda_{\rm FMW}=18$--180.  Applying this ratio to the
65~km DKIST wavelength gives
$\lambda_{\rm FMW}\simeq0.36$--3.6~km.  For low-coronal densities
$n_i=10^9$--$10^{10}$~cm$^{-3}$, the ion inertial length is
$d_i\simeq2.3$--7.2~m, giving
$\lambda_{\rm FMW}/d_i\simeq50$--$1.6\times10^3$.  For representative
low-coronal parameters $T_i=10^6$~K and $B=80$~G, the thermal proton
gyroradius is
\begin{equation}
 \rho_i=\frac{v_{{\rm th},i}}{\Omega_{ci}}
 =\frac{\sqrt{2m_i k_{\rm B}T_i}}{eB}
 \simeq0.17~{\rm m},
\end{equation}
which gives
$\lambda_{\rm FMW}/\rho_i\simeq2.1\times10^3$--$2.1\times10^4$.
The inferred wavelengths therefore remain larger than both characteristic
ion kinetic scales, so for the same ion heating mechanism to be present as in the magnetosphere, the KHI should lead to higher $k$-number modes. Recent multi-spacecraft MMS observations have revealed whistler waves at the CME boundary generated by KHI driven reconnection suggesting this pathway may be possible \citep{nykyri2026}. The KHI driven kinetic waves are generated by two- and three-dimensional KH reconnection which form
non-Maxwellian ion distributions containing free energy for such waves
\citep{nykyri2006,ma2019,nykyriMaJohnson2021}.  Wave heating is thus a
plausible complementary channel, although it has not yet been detected above
the DKIST vortices.

\section{Discussion and Conclusions}\label{sec:discussion}

The calculation indicates that the newly resolved vortices contain a
substantial local shear reservoir.  With the stated density and geometry, one
characteristic vortex contains about $2.2\times10^{24}$ erg.  If current sheets
reach a weakly collisional layer, magnetospheric reconnection scalings imply
that 28--44\% of the magnetic energy entering the exhaust can appear as bulk
heat.  For the illustrative filling factor, quiet-Sun heating requires 5--8\%
of the shear reservoir to be converted into reconnecting magnetic free energy
and transmitted upward.  Coronal holes require 14--21\%.  The same
normalization does not supply the active-region loss.

That active-region result has a narrow meaning.  It rules out the adopted
quiet-network normalization as a steady local heat source for active-region
radiative losses; it does not rule out KH-related dynamics in active regions.
There the larger reservoir is likely the magnetic free energy and helicity
stored by sustained twisting, shearing, and emergence of the guide-field
system.  KH vortices and other photospheric motions can help build, braid, or
thin current layers in such a stressed field.  Intermittent three-dimensional
reconnection, including paired sites away from the shear plane and
out-of-plane anti-parallel components where the guide field reverses locally,
can then release or redistribute the stored energy.  If the twisted structure
expands into the tenuous low corona and loses equilibrium, the same storage
and release path becomes eruptive, producing a flux rope and CME-like event
rather than the quasi-steady quiet-coronal heating estimated here.

A second result is the altitude ladder of Section~\ref{sec:ladder}.  Complete
conversion of the shear reservoir is self-limiting at the 184 G
marginal-stability field, equivalent to a $7.6^\circ$ footpoint twist, and
stores 135 J m$^{-3}$ in layers that thin toward ion and electron inertial
widths.  Because the energy per unit sheet area at the ion width is nearly
height independent while the density drops by more than four orders of
magnitude, dissipation of the same footpoint-driven sheets yields
$\approx$22 eV per ion at the photosphere (diluted to $\sim$14 K by
neutrals), $\sim10^{2}$ eV in the chromosphere, and $\approx$1--1.4 keV
($\sim10^{7}$ K) in the low corona.  The same footpoint twist does not make the
photosphere hot, because the released energy is shared with many neutrals.  It
can produce nanoflare-class exhaust temperatures only after the density has
fallen enough and provided the twist or associated current system survives
transit through the chromosphere.

The largest uncertainty is  the connection between the
photospheric source and the low corona.  The 500 km vertical extent in MURaM is
mostly below the visible surface.  It does not demonstrate propagation through
the chromosphere.  

The proposed sequence in this paper can be tested in the existing MURaM cube and also by generating new simulations with finite magnetic field tilts.  A useful
analysis would follow the transverse magnetic field, current-sheet thickness,
Ohmic and ambipolar heating, vertical Poynting
flux, and field-line connectivity around each identified KH interface.  KH
growth should precede magnetic compression and localized current formation.
The current sheets should be short, intermittent, and locally perpendicular to
the dominant guide field rather than one continuous vertical X-line.  Paired
connectivity changes would be evidence for the three-dimensional route in
Figure~\ref{fig:KHI}(d).  If these signatures are absent, the same analysis
will place an upper bound on $\xi_B\eta_{\rm up}$.

Coordinated DKIST spectropolarimetry and chromospheric and coronal diagnostics
could then search above the same flux boundary for bidirectional Doppler flows,
nonthermal line broadening, intermittent brightenings, and upward wave power.
Until such measurements are available, photospheric KHI should be regarded as
a possible contributor to quiet-coronal heating rather than a demonstrated
solution of the coronal-heating problem.

\begin{acknowledgments}
K.N. acknowledges support from NASA Living With a Star grant 80NSSC23K0899
and National Science Foundation grant 2308853.
\end{acknowledgments}

\section*{Data Availability}

This Letter introduces no new observational data.  The DKIST and MURaM
quantities are from \citet{kuridze2026}.  The calculation scripts (Python and
MATLAB) that reproduce all derived quantities, together with their complete
numerical output, are archived on Figshare at
doi:\href{https://doi.org/10.6084/m9.figshare.33236022}%
{10.6084/m9.figshare.33236022} (DOI activated upon publication).

\facilities{DKIST}

\software{MURaM \citep{kuridze2026}, NumPy, Matplotlib}

\end{document}